# Expected Message Delivery Time for Small-world Networks in the Continuum Limit

Hazer Inaltekin, Mung Chiang and H. Vincent Poor

Department of Electrical Engineering, Princeton University, Princeton, NJ 08544

{hinaltek, chiangm, poor}@princeton.edu

*Abstract*—**Small-world networks are networks in which the graphical diameter of the network is as small as the diameter of random graphs but whose nodes are highly clustered when compared with the ones in a random graph. Examples of small-world networks abound in sociology, biology, neuroscience and physics as well as in human-made networks. This paper analyzes the average delivery time of messages in dense small-world networks constructed on a plane. Iterative equations for the average message delivery time in these networks are provided for the situation in which nodes employ a simple greedy geographic routing algorithm. It is shown that two network nodes communicate with each other only through their short-range contacts, and that the average message delivery time rises linearly if the separation between them is small. On the other hand, if their separation increases, the average message delivery time rapidly saturates to a constant value and stays almost the same for all large values of their separation.**

## I. INTRODUCTION

Many networks arising in science [1], technology [2] and society [3] exhibit complex connections by means of small numbers of long-range links, and are surprisingly closely connected despite their extent. Such networks have become known as small-world networks. The most striking characteristic of small-world networks is their small graphical diameter even though they are highly clustered. One important field of science in which small-world network structures frequently arise is the theory of social networks where the connections among people in a social network are studied [4]. Social networks will be the major focus of this paper; however, the results obtained for the average delivery time of messages are also valid in similar small-world network models appearing in other fields of science and technology.

Papers such as [5] and [6] are experimental verifications of the small-world phenomenon in social networks. Watts and Strogatz proposed an elegant small-world network model in [7] which enables mathematical analysis of small-world networks. Algorithmic perspectives are investigated in [9], [10] and [11]. In this work, we also focus on the *algorithmic perspectives* of small-world networks, and analyze the average message delivery time when nodes run a greedy geographic forwarding algorithm to deliver a message to the final destination. We obtain an exact recursive expression for the average message delivery time in a small-world network rather than providing bounds as in [10] and [11].

The research was supported by the U.S. National Science Foundation under Grants ANI-03-38807 and CNS-06-25637.

## II. NETWORK MODEL

To investigate the average message delivery time in small-world networks, we consider an $R$ by $R$ square as our network domain. Let $\mathcal{D}$ represent this network domain, and $d(x, y)$ be distance between any two points $x$ and $y$ belonging to $\mathcal{D}$. We distribute $n$ *relay* nodes randomly (from a uniform distribution) over the network domain. Locations of *source* $s$ and *target* $t$ nodes will be assumed to be arbitrary so that we may analyze the average delivery time of a message as a function of their separation. We let $X_s$ be the location of the source node, $X_t$ be the location of the target node and $\{X_i\}_{i=1}^n$ be locations of $n$ relay nodes.

For each $n \geq 1$, we define $\mathcal{H}_n = \{X_i\}_{i=1}^n$. The random set $\mathcal{H}_n$ will be called the *node-location process*. We define the continuum limit as $\mathcal{H}_\infty = \bigcup_{n=1}^\infty \mathcal{H}_n$. An important topological property regarding $\mathcal{H}_\infty$ is the following.

*Theorem 1:* With probability one (w.p.1), $\mathcal{H}_\infty$ is a dense subset of the network domain.

*Proof:* Let $(\Omega, \mathcal{S}, \mathbb{P})$ be the underlying probability space on which all of the random variables are defined. Let $\mathbb{Q}$ be the set of rational numbers, and $\mathbb{Q}^2 = \mathbb{Q} \times \mathbb{Q}$. Let $\mathcal{T} = \mathcal{D} \bigcap \mathbb{Q}^2$. Choose any $x \in \mathcal{T}$. For any small ball $\mathcal{B}\left(x, \frac{1}{p}\right) \subseteq \mathcal{D}$ around $x$ with radius $\frac{1}{p}$ and $p$ being an integer greater than or equal to 1,[1] let $Z_n = \sum_{i=1}^n \xi_i$, and $\xi_i = \mathbb{1}_{\{X_i \in \mathcal{B}(x, \frac{1}{p})\}}$. Since, under our model, the $\xi_i$'s are independent and identically distributed (i.i.d) random variables with finite means $\mu > 0$, we have $\frac{Z_n}{n} \to \mu$ w.p.1. Observe that $Z_\infty = \left|\mathcal{H}_\infty \bigcap \mathcal{B}\left(x, \frac{1}{p}\right)\right| \geq \left|\mathcal{H}_n \bigcap \mathcal{B}\left(x, \frac{1}{p}\right)\right| = Z_n$ for all $n \geq 1$. Thus, $Z_\infty \geq \lim_{n\to\infty} Z_n = \infty$. As a result, there will be infinitely many points of $\mathcal{H}_\infty$ lying in $\mathcal{B}\left(x, \frac{1}{p}\right)$ w.p.1. Let $\Omega_{x,p} \subseteq \Omega$ be the set on which $\mathcal{H}_\infty$ has infinitely many points in $\mathcal{B}\left(x, \frac{1}{p}\right)$. We define $\Omega_0$ as $\Omega_0 = \bigcap_{x \in \mathcal{T}} \bigcap_{p=1}^\infty \Omega_{x,p}$. $\mathbb{P}\{\Omega_0\} = 1$ since it is an intersection of countably many sets having probability measure one. Take an $\omega \in \Omega_0$. Now, for any $x \in \mathcal{D}$ and $\epsilon > 0$, we can find $z \in \mathcal{T}$ and $p \geq 1$ such that $\mathcal{B}\left(z, \frac{1}{p}\right) \subseteq \mathcal{B}(x, \epsilon)$. Therefore, $\mathcal{H}_\infty(\omega) \bigcap \mathcal{B}(x, \epsilon) = \infty$. This completes the proof. ∎

### A. Connectivity Properties of Nodes

We assume that each node in the network maintains both local and long-range contacts as in other existing small-world

---

[1] $\mathcal{B}\left(x, \frac{1}{p}\right) = \{y \in \mathcal{D} : d(x, y) < \frac{1}{p}\}$

network models. We first describe how the set of local contacts of a node is formed.

**(i) Local Contacts of a Node:** All nodes in the network have a uniform communication range $r$. Nodes lying inside this communication range of a node form its local contacts. Let us consider a particular node $i$ with location $X_i$. Then, its *local-neighborhood* is the ball $\mathcal{B}(X_i, r)$.

**(ii) Long-range Contacts of a Node:** In addition to its local contacts, each node maintains some number of long-range links for communication. We assume that there are two types of long-range links: **(i)** *incoming long-range links* and **(ii)** *outgoing long-range links*. A node can receive a message through an incoming long-range link but cannot send any data over this link. On the other hand, a node can send a message over an outgoing long-range link but cannot receive anything. Such a differentiation between incoming and outgoing long-range links helps in simplifying some of the mathematical details.

### B. Limiting Process for Forming the Outgoing Long-range Contacts

We now put forward the rules for the limiting process for forming outgoing long-range contacts as the number of relay nodes goes to infinity.

1) *Rules for forming outgoing long-range contacts:* Source and target nodes are only allowed to choose their outgoing long-range contacts among relay nodes. Relay nodes are also only allowed to choose their outgoing long-range contacts among relay nodes. A node chooses the node at the receiving end of an outgoing long-range link at random uniformly among all other nodes which do not lie in its local neighborhood. After a node chooses a long-range contact, it is not allowed to change it as new relay nodes are added to the network.

2) *Initialization and the limiting process:* We initialize the number of relay nodes to $m$ where $m$ is an arbitrary integer greater than 1, and uniformly distribute $m$ relay nodes over $\mathcal{D}$. We place source and target nodes at arbitrary positions in $\mathcal{D}$. Nodes then choose their outgoing long-range contacts as explained above. We send the number of relay nodes to infinity by adding new relay nodes. As a new relay node is added, nodes which have not been able to pick an outgoing long-range contact yet are given a chance to form a long-range outgoing contact with this new relay node. Nodes which have already chosen their outgoing long-range contacts are not allowed to change them.

Let $E_{n,i}$ be the event that node $i \in \{s, t\} \bigcup \{1, 2, \cdots, n\}$ has an outgoing long-range contact when there are $n$ relay nodes located in $\mathcal{D}$. On this event, node $i$'s outgoing long-range contact will be uniformly distributed over $\mathcal{D} - \mathcal{B}(X_i, r)$. Therefore, the probability that node $i$ has an outgoing long-range contact $j$ ($i \xrightarrow{\text{LRC}} j$) in $\mathcal{A} \subseteq \mathcal{D}$ conditioned on $X_i$ and $E_{n,i}$ can be written as

$$\mathbb{P}\{\exists j \text{ in } \mathcal{A} \bigcap \mathcal{H}_n \text{ s.t } i \xrightarrow{\text{LRC}} j | X_i, E_{n,i}\} = \frac{\int_{\mathcal{A}} \mathbb{1}_{\{z \in \mathcal{A} - \mathcal{B}(X_i, r)\}} \mathrm{d}z}{\int_{\mathcal{D}} \mathbb{1}_{\{z \in \mathcal{D} - \mathcal{B}(X_i, r)\}} \mathrm{d}z}.$$

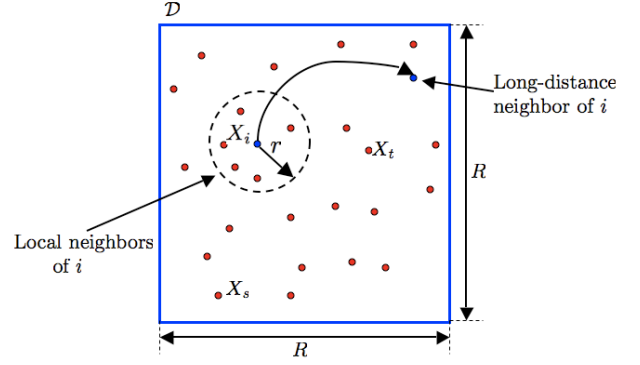

Fig. 1.   A typical realization of the network.

We have the following theorem when there are infinitely many relay nodes located in $\mathcal{D}$.

*Theorem 2:* For a given node $i \in \{s, t\} \bigcup \{1, 2, 3, \cdots\}$ and a given subset $\mathcal{A}$ of $\mathcal{D}$,

$$\mathbb{P}\{\exists j \text{ in } \mathcal{A} \bigcap \mathcal{H}_\infty \text{ s.t } i \xrightarrow{\text{LRC}} j | X_i\} = \frac{\int_{\mathcal{A}} \mathbb{1}_{\{z \in \mathcal{A} - \mathcal{B}(X_i, r)\}} \mathrm{d}z}{\int_{\mathcal{D}} \mathbb{1}_{\{z \in \mathcal{D} - \mathcal{B}(X_i, r)\}} \mathrm{d}z}.$$

*Proof:* Let $E_{n,i}$ be defined as above, and $A_{n,i} = \{\exists j \text{ in } \mathcal{A} \bigcap \mathcal{H}_n \text{ s.t } i \xrightarrow{\text{LRC}} j\}$. $\{E_{n,i}\}_{n \geq m}$ is an increasing sequence of events since nodes are not allowed to change their outgoing long-range contacts once they choose them. Therefore, whenever a node has an outgoing long-range contact for some $n \geq m$, it also has an outgoing long-range contact for all $k \geq n$. Similarly, $\{A_{n,i}\}_{n \geq m}$ is an increasing sequence of events. Let $E_{\infty,i} = \bigcup_{n \geq m} E_{n,i}$ and $A_{\infty,i} = \bigcup_{n \geq m} A_{n,i}$. Since $E_{\infty,i}$ is a dense subset of $\mathcal{D}$, we have $\mathbb{P}(E_{\infty,i}) = \lim_{n \to \infty} \mathbb{P}(E_{n,i}) = 1$. Observe also that $A_{\infty,i} = \{\exists j \text{ in } \mathcal{A} \bigcap \mathcal{H}_\infty \text{ s.t } i \xrightarrow{\text{LRC}} j\}$. Then, by using the continuity of measure from below [12] and Bayes theorem, we have

$$\begin{aligned}
\mathbb{P}(A_{\infty,i}) &= \lim_{n \to \infty} \mathbb{P}(A_{n,i} | E_{n,i}) \mathbb{P}(E_{n,i}) \\
&= \frac{\int_{\mathcal{A}} \mathbb{1}_{\{z \in \mathcal{A} - \mathcal{B}(X_i, r)\}} \mathrm{d}z}{\int_{\mathcal{D}} \mathbb{1}_{\{z \in \mathcal{D} - \mathcal{B}(X_i, r)\}} \mathrm{d}z}. \qquad \blacksquare
\end{aligned}$$

A typical realization of the network, and the probability with which nodes form their outgoing long-range contacts are depicted in Fig. **??** and Fig. 2, respectively.

### C. $\delta$-Greedy Geographic Forwarding Rule

Nodes use a simple $\delta$-greedy geographic forwarding ($\delta$-GGF) rule to deliver their messages to their intended target nodes. In $\delta$-GGF, the only global information needed to deliver a message is the location of its final destination, which can be encoded inside the message by the message originator.

According to this rule, any message holder having distance $d \in [kr, (k+1)r)$ to $X_t$ for some $k \geq 1$ chooses one of its local contacts lying in $\mathcal{B}(X_t, kr)$ and providing a forward progress at least $r - \delta$ in the direction of the target node for some $\delta > 0$ as a next hop local contact. After a node chooses its next hop local contact, it is not allowed to change it as new relay nodes are added to the network even if there are some other local contacts providing forward progress larger than the chosen one. This rule can be interpreted as that nodes have *bounded greediness* while choosing their next hop local contacts.

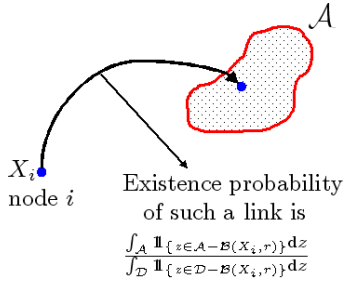

Fig. 2. An illustration of the probability with which a node chooses its outgoing long-range contacts.

If the current message holder has an outgoing long-range contact providing forward progress in the direction of the target node larger than the progress provided by the local contact chosen as described above, this long-range contact is preferred over the local contact.

## III. MESSAGE TRAJECTORY AND AVERAGE MESSAGE DELIVERY TIME

### A. Message Trajectory

In accordance with the search algorithm, a message wanders around the network hop by hop until it reaches the target. When there are $n$ relay nodes, we denote the resulting random trajectory of the message by $M_k^{(n)}$, where $M_k^{(n)}$ is the location of the message at the $k$th step. $M_0^{(n)}$ is always equal to $X_s$. We define the message delivery time $\tau_n$ under $\mathcal{H}_n$ to be equal to the first time the message reaches the target node. Therefore, $\tau_n$ can be written as $\tau_n = \inf\{k \geq 0 : M_k^{(n)} = X_t\}$. Note that $\tau_n = \infty$ if the forwarding rule cannot find a path from the source node to the target node. $\tau_n$ also depends on the locations of source and target nodes since the message starts its journey at $X_s$ and finishes it at $X_t$. However, we will not explicitly write it as a function of $X_s$ and $X_t$ by keeping this fact in mind.

*Definition 1:* We say a forwarding rule is *loop-free* if it does not form any loop while delivering the message to the target node.

*Definition 2:* We say a forwarding rule is *convergent* if for any given $X_s$ and $X_t$, $\tau_n$ converges to a limiting real-valued random variable w.p.1 as $n$ goes to infinity under the rules in II-B for forming long-range contacts.

For example, the shortest path routing is a convergent forwarding rule since the message delivery time monotonically decreases as new relay nodes added to the network domain. For convergent forwarding rules, we let $\lim_{n\to\infty} \tau_n = \tau$ to be the message delivery time under $\mathcal{H}_\infty$.

Let $\mathcal{F}$ denote the collection of all *convergent* and *loop-free* forwarding rules. We have the following theorem establishing the relationship between $\mathbb{E}[\tau_n]$ and $\mathbb{E}[\tau]$ for this type of rules.

*Theorem 3:* Fix $X_s$ and $X_t$. Then, for any forwarding rule in $\mathcal{F}$ with $\mathbb{P}\{\tau_n > B\} = o(n^{-1})$ for some finite $B > 0$,

$$\lim_{n\to\infty} \mathbb{E}[\tau_n \mathbb{1}_{\{\tau_n < \infty\}}] = \mathbb{E}[\tau]. \tag{1}$$

*Proof:* Since the forwarding rule is loop-free, we have

$$\tau_n \mathbb{1}_{\{\tau_n < \infty\}} = \tau_n \mathbb{1}_{\{\tau_n \leq B\}} + \tau_n \mathbb{1}_{\{B < \tau_n \leq n+1\}}.$$

Since $\tau_n$ converges to $\tau$ w.p.1, we have $\mathbb{P}\{\tau > B\} = 0$. Therefore, by using bounded convergence theorem, we have $\lim_{n\to\infty} \mathbb{E}[\tau_n \mathbb{1}_{\{\tau_n \leq B\}}] = \mathbb{E}[\tau]$. By using the condition on the decay rate of the tail of the distribution of $\tau_n$, we have

$$0 \leq \lim_{n\to\infty} \mathbb{E}[\tau_n \mathbb{1}_{\{B < \tau_n \leq n+1\}}] \leq \lim_{n\to\infty} (n+1)\mathbb{P}\{\tau_n > B\}$$
$$= 0.$$

As a result, $\lim_{n\to\infty} \mathbb{E}[\tau_n \mathbb{1}_{\{\tau_n < \infty\}}] = \mathbb{E}[\tau]$. ∎

We now show that Theorem 3 holds for $\delta$-GGF. To this end, we need to show a series of lemmas.

*Lemma 1:* $\delta$-GGF is a convergent forwarding rule for any $\delta > 0$.

*Proof:* We prove this lemma by induction. Consider the event $\Omega_0$ on which $\mathcal{H}_\infty$ is a dense subset of $\mathcal{D}$, and let $d = d(X_s, X_t)$. $\tau_n = 0$ if $d = 0$, and $\tau_n = 1$ if $0 < d < r$. Therefore, it is convergent for $0 \leq d < r$. Take any $\omega \in \Omega_0$. Since $\mathcal{H}_\infty(\omega)$ is a dense subset of $\mathcal{D}$, the message eventually enters $\mathcal{B}(X_t, r)$, and therefore $\tau_n(\omega)$ converges to 2 for any $\omega \in \Omega_0$ and for all $d \in [r, 2r)$. Similarly, it can be shown that $\tau_n$ either converges 2 or 3 on $\Omega_0$ for all $d \in [2r, 3r)$. Now, assume that $\tau_n$ converges on $\Omega_0$ for all $k \leq K$ and $kr \leq d < (k+1)r$. We will show that it is also correct for any given $d \in [(K+1)r, (K+2)r)$. Take any $\omega \in \Omega_0$. Note that $\mathcal{B}(X_s, r) \bigcap \mathcal{B}(X_t, (K+1)r)$ has positive area. Since $\mathcal{H}_\infty(\omega)$ is a dense subset of $\mathcal{D}$, $M_1^{(n)}(\omega)$ will eventually lie in $\mathcal{B}(X_t, (K+1)r)$ through either a short-range contact or a long-range contact for all $n$ large enough. Since the next hop short-range and long-range contacts are fixed once they are chosen, the relay node holding the message after the first transmission will be the same for all $n$ large enough. Now, this relay node located at $M_1^{(n)}(\omega)$ and currently holding the message can be interpreted as the new source node of the message. By induction hypothesis, the message delivery time starting at this relay node converges. Therefore, $\tau_n$ also converges on $\Omega_0$ for all $d \in [(K+1)r, (K+2)r)$. ∎

We write $f_1(k) \sim f_2(k)$ as $k \to \infty$ if $\lim_{k\to\infty} \frac{f_1(k)}{f_2(k)} = 1$.

*Lemma 2:* Let $a_k \to 0$ and $b_k \to \infty$ as $k \to \infty$. Then, $(1+a_k)^{b_k} \sim \exp(a_k b_k)$ if and only if $(a_k)^2 b_k \to 0$ as $k \to \infty$.

*Lemma 3:* For any given source-target separation $d = d(X_s, X_t)$ and $\delta > 0$, there exists a constant $B > 0$ such that $\tau_n$ under $\delta$-GGF satisfies $\mathbb{P}\{\tau_n > B\} \leq \exp(-c \cdot f(n))$, where $c > 0$ is a constant independent of $d$ and $\delta$, and $f(n) \sim n^{2\epsilon}$ for some $\epsilon \in (0, \frac{1}{4})$.

*Proof:* Choose an $\epsilon$ belonging to $\left(0, \frac{1}{4}\right)$. Divide $\mathcal{D}$ into small sub-squares with side length $\epsilon_n = \frac{R}{\lceil n^{\frac{1}{2}-\epsilon} \rceil}$. Let $h(n) =$

$\left\lceil n^{\frac{1}{2}-\epsilon} \right\rceil^2$ be the number of small sub-squares. Index sub-squares by $j \in \{1, 2, \cdots, h(n)\}$. If $S_j$ represents the $j$th sub-square, we have $\mathcal{D} = \bigcup_{j=1}^{h(n)} S_j$. Index relay nodes by $i \in \{1, 2, \cdots, n\}$. Then, $\mathbb{P}\{X_i \in S_j\} = \left\lceil n^{\frac{1}{2}-\epsilon} \right\rceil^{-2}$. Thus,

$$\mathbb{P}\{X_i \notin S_j, i \in \{1, 2, \cdots, n\}\} = \left(1 - \left\lceil n^{\frac{1}{2}-\epsilon} \right\rceil^{-2}\right)^n. \quad (2)$$

By using Lemma 2, we have

$$\begin{aligned}\mathbb{P}\{X_i \notin S_j, i \in \{1, 2, \cdots, n\}\} &\sim \exp\left(-n \left\lceil n^{\frac{1}{2}-\epsilon} \right\rceil^{-2}\right) \\ &\leq \exp\left(-0.25 \cdot n^{2\epsilon}\right).\end{aligned}$$

Let $E_n = \{\omega \in \Omega : \exists j \in \{1, 2, \cdots, h(n)\} \text{ s.t. } X_i(\omega) \notin S_j, \forall i \in \{1, 2, \cdots, n\}\}$. By using the union bound, we have

$$\begin{aligned}\mathbb{P}(E_n) &\leq h(n) \left(1 - \left\lceil n^{\frac{1}{2}-\epsilon} \right\rceil^{-2}\right)^n \\ &= \exp\left(-c.f(n)\right), \quad (3)\end{aligned}$$

where $c = 0.25$ and $f(n) = n^{2\epsilon} - 4\log(2) - 4(1-2\epsilon)\log(n)$.

On the complement $E_n'$ of the event $E_n$, the message can make forward progress at least $r - \delta$ amount towards destination at each hop for all $n$ large enough. Let $B = \left\lfloor \frac{d}{r-\delta} \right\rfloor + 1$. Then, $E_n' \subseteq \{\tau_n \leq B\}$. Thus,

$$\begin{aligned}\mathbb{P}\{\tau_n > B\} &\leq 1 - \mathbb{P}(E_n') \\ &\leq \exp\left(-c \cdot f(n)\right). \quad \blacksquare\end{aligned}$$

*Theorem 4:* If nodes use the $\delta$-GGF rule to relay messages, then $\tau_n$ converges to a limiting random variable $\tau$ and $\lim_{n \to \infty} \mathbb{E}\left[\tau_n \mathbb{1}_{\{\tau_n < \infty\}}\right] = \mathbb{E}\left[\tau\right]$.

Theorem 4 allows us to relate the average message delivery time for dense networks with finitely many nodes to the average message delivery time in the continuum limit for the $\delta$-GGF rule. Therefore, we will focus on the continuum limit and the $\delta$-GGF rule for the rest of the paper.

### B. Average Message Delivery Time

We now calculate the average message delivery time in small-world networks in the continuum limit when relay nodes are uniformly distributed over $\mathcal{D}$. We let $d = d(X_s, X_t)$ be the distance between source and target nodes. In order to eliminate possible edge effects that can occur, we place the target node at a position $X_t$ satisfying $\mathcal{B}(X_t, d+r) \subset \mathcal{D}$. This assumption, in turn, necessitates $R$ be greater than $2 \cdot (d+r)$. One can also extend the analysis presented below to small-world networks constructed on the surface of a sphere and on the surface of a torus where we do not run into edge-effect problem. Due to space limitations, we focus here only on the rectangular network domains.

Consider the $\delta$-GGF rule for some $\delta > 0$, and concentric cylinders $\mathcal{C}(X_t, kr, (k+1)r) = \mathcal{B}(X_t, (k+1)r) - \mathcal{B}(X_t, kr)$, $k \geq 0$, centered at $X_t$ with thickness $r$. In the continuum limit, it is not hard to see that the progress of a message toward the destination node at each hop is at least $r - \delta$ units, and the message enters to an inner concentric cylinder from an outer one due to $\mathcal{H}_\infty$ being dense in $\mathcal{D}$. As a result, if nodes do not have any long-range contacts, it can be shown that $\tau$ satisfies the following theorem.

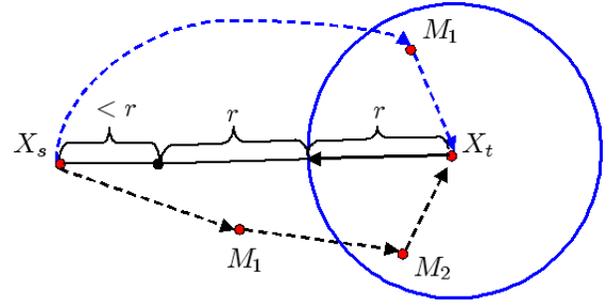

Fig. 3. Exemplary trajectories of the message when $2r \leq d(X_s, X_t) < 3r$.

*Theorem 5:* Under $\delta$-GGF with no long-range contacts, $\tau$, with probability one, is equal to

$$\tau = \begin{cases} 0 & \text{if } X_s = X_t, \\ \left\lfloor \frac{d(X_s, X_t)}{r} \right\rfloor + 1 & \text{if } d(X_s, X_t) > 0. \end{cases} \quad (4)$$

Let us now turn our attention to the more interesting case in which nodes maintain long-range contacts in addition to those with their local neighbors. Consider any two source nodes located on the boundary $\partial \mathcal{B}(X_t, d)$ of $\mathcal{B}(X_t, d)$. Due to the symmetry of the problem, neither of the nodes is better than the other in delivering the message to the destination. Therefore, the expected value of $\tau$ is only a function of the separation between the source-target pair, and there is no ambiguity in the following definition.

*Definition 3:* $g(d) \triangleq \mathbb{E}^d[\tau]$ is the expected delivery time of a packet to the final destination $X_t$ starting from the source node $X_s$ when $d(X_s, X_t) = d$.

Let us first calculate $g(d)$ for several ranges of $d$ before going into a more general solution. $g(0) = 0$ since there is no need to transmit the message. For $0 < d < r$, $g(d) = 1$ since source and target nodes can directly communicate with each other. For $r \leq d < 2r$, $g(d) = 2$ since the source can reach the target in exactly two hops due to $\mathcal{H}_\infty$ being dense in $\mathcal{D}$. For $2r \leq d < 3r$, $g(d)$ is a little tricker to obtain. The key is to analyze where the message can be located after the first hop. It can be delivered to $X_t$ in two hops if it makes its way through $\mathcal{B}(X_t, r)$ at the first hop. Otherwise, it first enters to the disc $\mathcal{C}(X_t, r, 2r)$, and then it is delivered to the target in two hops thereafter. As a result, the second path takes 3 hops to deliver the message to the final destination. Exemplary trajectories of the packet are shown in Fig. 3.

These observations lead to $g(d) = 1 + \mathbb{P}\{0 \leq d(M_1, X_t) < r\} + 2 \cdot \mathbb{P}\{r \leq d(M_1, X_t) < 2r\}$, where $M_k$, $k \geq 0$, represents the location of the message at the $k$th hop under $\mathcal{H}_\infty$. By letting $\alpha = \mathbb{P}\{0 \leq d(M_1, X_t) < r\} = \frac{\pi r^2}{R^2 - \pi r^2}$, we obtain $g(d) = 1 + \alpha + 2(1 - \alpha)$ for $2r \leq d < 3r$.

General calculations for any $d$ lying in $\left[k \cdot r, (k+1) \cdot r\right)$, $k \geq 2$, are also in this spirit. We first look at where the message can be located at the first hop, then analyze the expected delivery time from this point on. This analysis leads to a recursive solution for $g(d)$ for any value of $d$. To this end, we let $g(d) = g_0$ when $d = 0$, $g(d) = g_1$ when $0 < d < r$, $g(d) = g_k$ when $(k-1) \cdot r \le d < k \cdot r$, $2 \le k \le \lfloor \frac{R}{2r} - 1 \rfloor$ and $g(d) = g_{\lfloor \frac{R}{2r} - 1 \rfloor + 1}$ when $\lfloor \frac{R}{2r} - 1 \rfloor r \le d < \frac{R}{2} - r$.

Consider $g_{k+1}$ when $2 \le k \le \lfloor \frac{R}{2r} - 1 \rfloor$. Then,

$$
\begin{aligned}
g_{k+1} &= 1 + \mathbb{E}\left[ \sum_{i=1}^{k} g_i \cdot \mathbb{1}_{\{(i-1) \cdot r \le d(M_1, X_t) < i \cdot r\}} \right] \\
&= 1 + \left(1 - \alpha(k-1)^2\right) g_k + \alpha \cdot \sum_{i=1}^{k-1} (2i-1) g_i.
\end{aligned}
$$

Note that in the first equality above, there is no loss of generality in writing the limits of the indicator function as $0 \le d(M_1, X_t) < r$ when $i = 1$ since $\mathbb{P}\{d(M_1, X_t) = 0\} = 0$. To obtain a second order non-constant coefficient linear recursive equation, we subtract $g_k$ from $g_{k+1}$.

$$
g_{k+1} - g_k = (g_k - g_{k-1}) \cdot \left(1 - \alpha(k-1)^2\right). \tag{5}
$$

Let $u_k = g_{k+1} - g_k$ and $\beta_k = 1 - \alpha(k-1)^2$. Then, $u_k$ satisfies the following first order linear recursive equation.

$$
u_k = \beta_k \cdot u_{k-1} \text{ for } k \ge 1, \tag{6}
$$

with the initial condition $u_0 = 1$. Observe that $u_k = \prod_{i=1}^{k} \beta_i$ for $k \ge 1$. Then, the solution for (5) is obtained as

$$
g_{k+1} = 1 + \sum_{j=1}^{k} \prod_{i=1}^{j} \beta_i \text{ for } k \ge 1. \tag{7}
$$

The following theorem summarizes these findings.

*Theorem 6:* Consider a small-world network constructed on $\mathcal{D}$, and containing nodes that have local communication range $r$ and one uniformly distributed long-range outgoing contact on $\mathcal{D}$. If nodes employ the $\delta$-GGF rule to relay messages, then the average message delivery time, for $1 \le k \le \lfloor \frac{R}{2r} - 1 \rfloor$, is given by $g_{k+1} = 1 + \sum_{j=1}^{k} \prod_{i=1}^{j} \beta_i$, where $g_k = g(d) = \mathbb{E}^d[\tau]$ when $(k-1)r \le d < kr$, $g(d) = g_{\lfloor \frac{R}{2r} - 1 \rfloor + 1}$ when $\lfloor \frac{R}{2r} - 1 \rfloor r \le d \le \frac{R}{2} - r$, $\alpha = \frac{\pi r^2}{R^2 - \pi r^2}$, and $g_0 = 0$ and $g_1 = 1$ are the initial conditions.

In Fig 4-a and Fig. 4-b , we plotted the change of average message delivery time with respect to the separation between source and destination nodes for two different network sizes. In both figures, the horizontal axis is normalized with respect to the local communication range $r$ of nodes. Fig. 4 reveals that the average message delivery time increases linearly for small values of separation between a source and destination pair. On the other hand, it quickly converges to a constant, and remains essentially the same at this constant value for a broad range of values of source-destination separations. This means that messages are first forwarded by means of long-range contacts until they enter a certain range of the destination. From this point on, they are delivered to the destination through local contacts. This quantifies the observation of Travers and Milgram [5]: "*Chains which converge on the target principally by using geographic information reach his hometown or surrounding areas readily, but once there often circulate before entering targets circle of acquaintances.*".

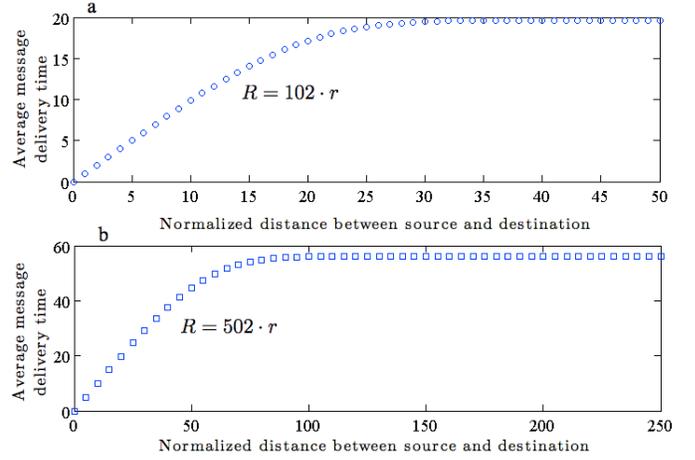

Fig. 4. Change of average message delivery time as a function of normalized source-destination separation. a, $R = 102 \cdot r$. b, $R = 502 \cdot r$. In both figures, the horizontal axis is normalized with respect to $r$.

## IV. Conclusions

Small-world networks arise in many disciplines of science including biology, neurology, sociology and computer science. In this work, we have focused on the average delivery time of messages to a final destination in dense small-world networks when nodes use a local geographic forwarding rule to relay messages. Existing work on small-world networks only provides bounds on this average message delivery time. On the other hand, in this paper, we have presented a technique based on the first-step analysis for calculating an exact formula of the packet delivery time in a small-world network constructed on a plane.